\documentclass[english]{article}
\usepackage[T1]{fontenc}
\usepackage[latin1]{inputenc}
\usepackage{geometry}
\makeatletter

\providecommand{\tabularnewline}{\\}

\newcommand{\lyxaddress}[1]{
\par {\raggedright #1
\vspace{1.4em}
\noindent\par}
}

\usepackage{babel}
\makeatother
\begin{document}

\title{Comment on the statistical analysis in \char`\"{}A new experimental
limit for the stability of the electron\char`\"{} by H.V. Klapdor-Kleingrothaus,
I.V. Krivosheina and I.V. Titkova}

\author{A.Derbin\textbf{$^{a}$}, A.Ianni\textbf{$^{b}$,} and O.Smirnov\textbf{$^{c}$.}}

\maketitle
$\:$

\lyxaddress{$a$)St. Petersburg Nuclear Physics Inst. - 188 350 Gatchina Petersburg
Reg., Russia; derbin@pnpi.spb.ru}

\lyxaddress{$b$)L.N.G.S. SS 17 bis Km 18+910, I-67010 Assergi(AQ), Italy; Aldo.Ianni@lngs.infn.it}

\lyxaddress{$c$)Joint Institute for Nuclear Research, Joliot Curie 6, 141980,
Dubna Moscow region, Russia; osmirnov@jinr.ru}

\begin{abstract}
We have revealed evident errors in the statistical analysis, performed
by Klapdor-Kleingrothaus et al in a recently published paper \cite{KKKT}
to establish a limit on the stability of electron with respect to
the decay into $\nu+\gamma$. The performed reestimation of the sensitivity
of the experimental setups to the 256 keV gamma emitted in the hypothetical
electron decay, has shown that the limits on the electron stability
and charge nonconservation parameter $\epsilon_{e\nu\gamma}^{2}$
presented in \cite{KKKT}, have been overestimated by at least a factor
of 5.

PACS: 13.35.-r; 13.40.Hq; 14.60.Cd; 11.30.-j

Keywords: Stability of the electron;  Electric charge conservation
\end{abstract}
Incongruity of the analysis is evident already from the comparison
of the sensitivity estimate with the \char`\"{}1$\sigma$\char`\"{}
method and the results of the maximum likelihood and $\chi^{2}$-
analysis, the latter two give the result by a factor of 3-5 better.
The most evident manifestation of errors in the analysis can bee seen
in Table 7 of Ref.\cite{KKKT} (called hereafter the KKKT article):

\begin{enumerate}
\item The best fit of the ANG2 data set contains 89.444$\pm$63.058 events%
\footnote{We are reproducing the number of significant digits, following the
original text of the KKKT article.%
} in the peak corresponding to $\sim$256 keV ($m_{e}/2$) $\gamma$
from hypothetical electron decay $e\rightarrow\nu+\gamma$ for the
ME case%
\footnote{See the original text of the KKKT article \cite{KKKT} to explain
the abbreviations, we are also citing a number of values from the
KKKT article without going too much into details.%
}. As follows from the text of the article the error in this number
is cited at 68\% c.l. (it is called an indication of a signal on a
1.4 $\sigma$ c.l.). The corresponding upper limit on the number of
events in the peak, $\lambda$, already at 50\% c.l., should be higher
than the central value of 89 in most practical cases of any almost-symmetrical
$\chi^{2}$- profile shapes. Instead of the above it is claimed to
be only 38 events at 68\% c.l.
\item The upper limit on the number of events in the 256 keV peak obtained
for ANG4 set, is 4.789 events at 68\% c.l. This value is lower than
1$\sigma$ statistical error on the content of a single bin in Fig.
8d, while the FWHM peak width is about 20 bins and the mean bin content
is $>50$ events. 
\end{enumerate}
These evident errors, together with the fact that an \char`\"{}indication
of a signal on 1.4 $\sigma$\char`\"{} for 1.94$\times10^{26}$ yr
is excluded both by the results of the Borexino \cite{Borexino} and
DAMA \cite{DAMA} collaborations with more than 90\% probability,
convinced us to have a closer look at the results of KKKT. The results
from the KKKT article are summarized in Table \ref{TableKKKT} (the
data are taken from Tables 5 and 7 of the KKKT article; for the sake
of simplicity we are citing only the ME case at 68\% c.l.). These
are the results of the analysis of experimental data with the 1$\sigma$
method (\char`\"{}$1\sigma$\char`\"{} column) and with the standard
least square procedure ({}``$\chi^{2}$'' column). One can see incompatibility
of the values in $N_{peak}$ and $\lambda$ subcolumns, the most evident
discrepancies are described above.

In order to check the achievable sensitivity when looking for the
gaussian shape on the linear background, we have applied a toy model
consisting of a gaussian peak superimposed on the linear background.
For the toy model we used the background level for the corresponding
data set from the KKKT article, the region of analysis was set to
100 keV with the bin width of 0.36 keV, and the 1$\sigma$ width of
the gaussian peak corresponded to the Doppler- broadened line of 3.25
keV. The number of the events that can be eliminated at 1$\sigma$
level was defined for a large number of samples by using the MC method.
For each sample a set of randomly distributed events was simulated
with the fixed mean number of events in the gaussian peak, then we
fitted it with linear+Gauss analytical shape. The sensitivity at 68\%
confidence level in this approach corresponds to the mean number of
events in the peak for which the $\chi^{2}$ value increases by $\Delta\chi^{2}=1$.
The results are presented in \char`\"{}MC\char`\"{} column of Table
\ref{TableOur}. 

Comparing the $\lambda$ values in MC column to the KKKT $1\sigma$
estimate one can see that the toy model gives the same level of sensitivity,
confirming the KKKT estimation with the $1\sigma$ method. Nevertheless,
the values of $\lambda$ obtained by KKKT with the $\chi^{2}$ method
are significantly (by 2-10 times) lower than their own estimates with
the 1$\sigma$ method. 

There is no description of the $\chi^{2}$- profile analysis in the
text of the KKKT article. It is also not clear how many free parameters
were used, what is the precise n.d.f, and whether the result depends
on the lower and upper limits of the analysis region. The number of
the events in the peak ({}``peak area'') seem to have been taken
directly from the minimization program together with the error on
this number. In principle, the correct limits with acceptable precision
can be reproduced by using these data. The results are presented in
Table \ref{TableOur}. The data of $\lambda$ are obtained assuming
the normal shape for the $\chi^{2}$- profile with the central value
and variation taken from the third column of Table 7 of the KKKT article
(these values are reproduced in $N_{peak}$ column of Table \ref{TableKKKT}).
Our limits have been calculated by using the Bayesian approach (see
i.e. \cite{PDG}, the prior knowledge in our case is the restriction
of the positively defined effect). The limits on the life-time recalculated
by using these values are presented in the next column. Only the ME
case for 68\% c.l. is shown, analysis for 90\% c.l. and the AE cases
can be performed in the similar way with the same conclusions. 

\begin{table}
\begin{centering}\begin{tabular}{cccccc}
\hline 
&
\multicolumn{2}{c}{1$\sigma$ method}&
\multicolumn{3}{c}{$\chi^{2}$}\tabularnewline
Detector&
 $\lambda$&
$\tau$&
$N_{peak}$&
$\lambda$&
$\tau$\tabularnewline
&
events&
years&
({}``peak area'')&
events&
years\tabularnewline
\hline 
ANG1&
49&
$5.8\times10^{24}$&
$-38.187\pm51.077$&
$13.216$&
$2.146\times10^{25}$\tabularnewline
\hline 
ANG2&
61&
$3.3\times10^{25}$&
$89.444\pm63.058$&
$38.354$&
$5.285\times10^{25}$\tabularnewline
\hline 
ANG3&
64&
$2.2\times10^{25}$&
$-38.301\pm67.374$&
$13.216$&
$10.76\times10^{25}$\tabularnewline
\hline 
ANG4&
46&
$2.0\times10^{25}$&
$-76.249\pm47.401$&
$4.789$&
$19.33\times10^{25}$\tabularnewline
\hline 
ANG5&
73&
$3.0\times10^{25}$&
$-33.273\pm75.947$&
$14.343$&
$15.69\times10^{25}$\tabularnewline
\hline
\end{tabular}\par\end{centering}

\caption{\label{TableKKKT}Data from the KKKT article (68\% c.l.; ME case
only).$\lambda$ is an upper limit on the number of events in the
peak from the hypothetical electron decay, $\tau$ is the corresponding
life-time limit.}
\end{table}

\begin{table}
\begin{centering}\begin{tabular}{cccc}
\hline 
&
MC&
\multicolumn{2}{c}{$\chi^{2}$}\tabularnewline
Detector&
$\lambda$(68\% c.l.)&
$\lambda$(68\% c.l.)&
$\tau$ (68\% c.l.)\tabularnewline
&
events&
events&
years\tabularnewline
\hline 
ANG1&
46&
36&
$7.8\times10^{24}$\tabularnewline
\hline 
ANG2&
58&
119&
$1.7\times10^{25}$\tabularnewline
\hline 
ANG3&
62&
51&
$2.8\times10^{25}$\tabularnewline
\hline 
ANG4&
44&
24&
$3.9\times10^{25}$\tabularnewline
\hline 
ANG5&
71&
61&
$3.7\times10^{25}$\tabularnewline
\hline
\end{tabular}\par\end{centering}

\caption{\label{TableOur}Reestimation of the life-time limits (68\% c.l.;
ME case only). }
\end{table}

As it is seen from our estimations the best limit that can be obtained
($\tau=3.9\times10^{25}$ yr at 68\% c.l. for ANG4 setup) is very
close to that already existed for the HPGe ($\tau=3.7\times10^{25}$
yr at 68\% c.l. \cite{HPGe}). The most surprising fact is that it
practically coincides with the estimate of the sensitivity obtained
by KKKT themselves with the $1\sigma$ method. As a result, the life-time
limits in the KKKT article from the $\chi^{2}$ analysis are by a
factor of 2 till 10 stronger than the estimated sensitivity, and by
a factor 3 till 5 higher than the values that can be obtained from
the values presented as the \char`\"{}Peak area\char`\"{}.%
\footnote{The same is true for the limits on the charge- nonconservation parameter
$\epsilon_{e\rightarrow\nu\gamma}^{2}$, derived from the upper limits
on the electron life-time. Using the best established limit on the
electron life-time ($\tau_{e\rightarrow\nu\gamma}>4.6\times10^{26}$
y, 68\% c.l. \cite{Borexino}) and the upper limit on the photon mass
($m_{\gamma}<7\times10^{-19}$ eV \cite{PDG}), from formula $\epsilon_{e\rightarrow\nu\gamma}^{2}=\left(\frac{m_{\gamma}}{m_{e}}\right)^{2}\frac{5.6\times10^{-25}}{\tau_{e\rightarrow\nu\gamma}[y]}$
\cite{DAMA} one can obtain the best restriction on the parameter
$\epsilon_{e\rightarrow\nu\gamma}^{2}<0.23\times10^{-98}$.%
}

In the proper analysis one should take the signal from all electrons
(the AE case) into account, not only from the outer shells, that can
obviously change the sensitivity. The limit for the AE case is inferior
to the existing for the HPGe, so there is no need in a more detailed
analysis. 

Some words should be written on the model used to describe the underlying
background. The quality of the fit for 4 sets is bad ($\chi^{2}\simeq390/280$,
this value of $\chi^{2}$ from the formal statistical point of view,
rejects the model with a very high probability) and only for the ANG2
set it has an acceptable quality ($\chi^{2}\simeq280/280$). If the
data are obtained under the same conditions (which seems to apply
at least to 4 detectors of Setup 1), then the model should give a
statistically compatible description for all sets. The quantitative
comparison of the data sets can be performed using Fischer's F-distribution
$\frac{\chi_{2}^{2}}{\chi_{1}^{2}}=F(p,\nu,\nu)$ as a significance
test, where $\nu$ is a number of the degrees of freedom and $p$
is a confidence level (see i.e.\cite{Wolberg}). Solving equation
$F(p,280,280)=390/280$ with respect to $p$, one obtains the statistical
probability of the data set with lower $\chi^{2}$: $p=0.003$. This
definitely points on the systematics problem with the data set ANG2%
\footnote{or, if we assume that the linear model of the underlying background
is valid, the situation is inverted, and in this case the data for
all detectors except ANG2 should have systematic problems. The question
what the real situation is should be addressed to the KKKT authors. %
}. 

The linearity of the background has not been justified in the KKKT
article, moreover, all the compatible sets contain the statistically
evident hole in the background just in the place where the effect
is searched for.

\section*{Conclusions}

Our analysis of the data presented by KKKT in \cite{KKKT} has shown
that the upper limit on the electron decay is overestimated by at
least a factor of 5. The statistical analysis of the KKKT contains
evident errors. Moreover, one of the presented data sets (containing
an \char`\"{}indication of a signal\char`\"{} on 1.4 $\sigma$) is
statistically inconsistent with 4 others, pointing out on possible
systematic problems with the experimental data. The model used in
\cite{KKKT} to fit the underlying background has the same problems
of statistical incompatibility.

The best limit that can be obtained by using the KKKT data is comparable
to the one established previously for the HPGe. The KKKT restriction
for charge nonconservation $\epsilon_{e\rightarrow\nu\gamma}^{2}<0.86\times10^{-98}$
is not valid either, since it is based on the overestimated electron
life-time. Instead of the above the restriction $\epsilon_{e\rightarrow\nu\gamma}^{2}<2.3\times10^{-99}$
at 90\% c.l. can be calculated from the modern best limit on the electron
life-time $\tau_{e\rightarrow\nu\gamma}>4.6\times10^{26}$ yr (90\%
c.l.) established by the Borexino collaboration.

\end{document}